\documentclass[aps,pre,reprint,superscriptaddress,amsmath,amssymb]{revtex4-2}

\usepackage[T1]{fontenc}
\usepackage{graphicx}
\usepackage{dcolumn}
\usepackage{bm}
\usepackage{hyperref}
\usepackage{newtxtext,newtxmath}
\usepackage{xcolor}

\newcommand{\aLR}{\alpha_\mathrm{LR}}

\begin{document}

\title{Neural-network quantum state study of the long-range antiferromagnetic Ising chain}

\author{Jicheol Kim}
\affiliation{Department of Physics and Photon Science, Gwangju Institute of Science and Technology, Gwangju 61005, Korea}

\author{Dongkyu Kim}
\affiliation{Department of Physics and Photon Science, Gwangju Institute of Science and Technology, Gwangju 61005, Korea}
\affiliation{Kakao Corporation, Pangyo, Gyeonggi 13529, Korea}

\author{Dong-Hee Kim}
\email{dongheekim@gist.ac.kr}
\affiliation{Department of Physics and Photon Science, Gwangju Institute of Science and Technology, Gwangju 61005, Korea}


\begin{abstract}
We investigate quantum phase transitions in the transverse field Ising chain with algebraically decaying long-range (LR) antiferromagnetic interactions using the variational Monte Carlo method with the restricted Boltzmann machine employed as a trial wave function ansatz. First, we measure the critical exponents and the central charge through the finite-size scaling analysis, verifying the contrasting observations in the previous tensor network studies. The correlation function exponent and the central charge deviate from the short-range (SR) Ising values at a small decay exponent $\alpha_\mathrm{LR}$, while the other critical exponents examined are very close to the SR Ising exponents regardless of $\alpha_\mathrm{LR}$ examined. However, in the further test of the critical Binder ratio, we find that the universal ratio of the SR limit does not hold for $\alpha_\mathrm{LR} < 2$, implying a deviation in the criticality. On the other hand, we find evidence of the conformal invariance breakdown in the conformal field theory (CFT) test of the correlation function. The deviation from the CFT description becomes more pronounced as $\alpha_\mathrm{LR}$ decreases, although a precise breakdown threshold is yet to be determined. 
\end{abstract}

\maketitle

\section{Introduction}
\label{sec:intro}

Artificial neural networks and machine learning have been influencing
the paradigm of physics research with a growing number of applications
on various subjects, including phase transitions
and critical phenomena in classical and quantum many-body systems
\cite{CarleoRMP,Carrasquilla2020,Carrasquilla2021,Dawid2022}.
In particular, the representation of a quantum wave function
by a neural network \cite{Carleo2017} provides an alternative
numerical platform combined with the variational Monte Carlo (VMC)
method to find the ground state of a many-body Hamiltonian. 
The neural-network quantum state (NQS) has extended its area of applications
to the Fermi and Bose Hubbard models \cite{Nomura2017,Saito2017},
real-time dynamics \cite{Carleo2017,Schmitt2020,Schmitt2022,Donatella2023},
open quantum systems \cite{Nagy2019,Hartmann2019,Vicentini2019,Yoshioka2019},
quantum state tomography \cite{Torlai2018,Torlai2019},
frustrated systems \cite{Nomura2021,Chen2022,Szabo2020,Westerhout2020,Ferrari2019,Choo2019,Liang2018},
and \textit{ab initio} simulations of molecules \cite{Choo2020,Pfau2020,Hermann2020}.
The NQS ansatz offers the high expressive capacity
often measured in terms of entanglement scaling
\cite{Deng2017,Chen2018,Glasser2018,Levine2019,Sharir2022}, 
proposing a complementary tool to conventional numerical
methods for studying quantum criticality.

In this paper, we investigate quantum phase transitions
in the transverse field Ising chain (TFIC) with algebraically decaying
long-range (LR) antiferromagnetic (AF) interactions
by employing the NQS ansatz for the VMC calculations.
LR-interacting quantum systems have attracted growing attention, 
both theoretical and experimental \cite{DefenuRMP}. 
The trapped-ion quantum simulation \cite{MonroeRMP} realized
the TFIC Hamiltonian with an LR interaction that maps to the form of
$1/r^{\aLR}$ with a tunable exponent $\aLR$, 
providing a controllable experimental platform to study quantum phase
transitions at and out of equilibrium \cite{Islam2011,Zhang2017,Li2023}.
The nearest-neighbor-interacting short-range (SR) TFIC is a textbook
example of quantum critical behavior in one dimension that belongs
to the universality class of the classical two-dimensional (2D)
Ising model \cite{SachdevBook}. 
However, such quantum-classical correspondence to the universality
of critical phenomena becomes nontrivial in the presence of LR interactions. 
A central question of how criticality depends on $\aLR$ is still
an active subject of various numerical and analytical studies
\cite{DefenuRMP,Defenu2020,Sak1973,Luijten2002,Angelini2014,Horita2017,Behan2017,Sandvik2010,Fey2019,Koziol2019,Humeniuk2020,Paulos2016,Dutta2001,Knap2013,Defenu2017,Langheld2022,Zhu2018,Shiratani2023,Ren2022,Koffel2012,Vodola2016,Sun2017,Puebla2019,Fey2016,Koziol2021,Kaicher2023,Halimeh2021,Vodola2014,Lepori2016}.

We revisit this question on the AF side of the LR interactions
for TFIC, where the breakdown of the Ising class in the critical ground
state seems to be very different from what is established
in the ferromagnetic (FM) counterpart
\cite{Dutta2001,Knap2013,Fey2016,Defenu2017,Langheld2022,Zhu2018,Puebla2019,Koziol2021,Shiratani2023}.
Because an exact solution is not available, constructing
the picture of how its criticality deviates from
the Ising class as $\aLR$ decreases relies primarily on
the collection of numerical observations.  
Despite various numerical studies characterizing the quantum
phase transition in AF-LR-TFIC at equilibrium
\cite{Koffel2012,Vodola2016,Fey2016,Sun2017,Puebla2019,Kaicher2023}
and out of equilibrium \cite{Puebla2019,Halimeh2021},
the picture remains incomplete in some parts,
which requires more numerical evidence for clarification.
Using the restricted Boltzmann machine (RBM) for the NQS ansatz \cite{Carleo2017},
we consider the moments of staggered magnetization
including the order parameter and the Binder ratio,
the two-point correlation function, 
and the entanglement entropy to examine the present picture
and improve the characterization of the phase transition
with increasing LR influences along the critical line.

We begin with brief reviews of previous results on
the characterization of the criticality.
The first study of AF-LR-TFIC \cite{Koffel2012} using
the time-dependent variational principle (TDVP)
found a phase transition for all $\alpha_\mathrm{LR} > 0$,
where it turned out that the critical exponent of the correlation
function decreases from the SR Ising value for $\aLR \lesssim 2$.
A significant increase in the central charge from the SR Ising value of $1/2$ 
was observed for $\aLR \lesssim 1$ in the TDVP \cite{Koffel2012}
and density matrix renormalization group (DMRG) \cite{Vodola2016}
calculations, based on which the breakdown of conformal invariance
was proposed \cite{Vodola2016}. 
However, more studies are needed because the central charge
is not a sufficient indicator of conformal invariance \cite{Patil2017}.
While we focus on the critical ground state, a violation of
the area law for the entanglement entropy was observed
in the offcritical area \cite{Koffel2012,Vodola2016,Kaicher2023}, 
and it was shown that the area law of the noncritical ground state
holds for $\aLR > 2$ \cite{Kuwahara2020}. 

On the other hand, contrasting evidence was found in the other 
DMRG calculations \cite{Sun2017,Puebla2019}, where the estimates of
the critical exponents $\nu \simeq 1$ and $\beta \simeq 1/8$ and
the dynamic exponent $z \simeq 1$ were in agreement with
the SR Ising values for all examined $\aLR$ between $0.4$ and $3$.
However, these DMRG estimates of the critical exponents have not
been fully verified in different approaches.
Linked cluster expansion calculations \cite{Fey2016} reported
$z\nu = 1.7(5)$ for $\aLR = 2$ while $z\nu \approx 1$ for $\aLR = 9/4$.
Previous quantum Monte Carlo (QMC) calculations with stochastic
series expansion \cite{Koziol2021} provided lower values of $\nu$ 
and $\beta$ in its examined range of $\aLR \ge 2$.
While partial disagreements exist between these previous estimates,
the TDVP and DMRG results together suggest an interesting possibility
that some of the exponents can still be very close to the SR Ising values
even for a small $\aLR$ where the central charge indicates a deviation,
raising the need for verification with a different numerical approach.

Apart from the question about criticality at a small $\aLR$,
another issue we want to address is the possibility of conformal invariance
breakdown that may occur below a certain value of $\aLR$.
The scenario of conformal invariance breakdown was also proposed
in the study of the Kitaev chain with LR pairing \cite{Vodola2014,Lepori2016}
which becomes equivalent to the Ising chain in the SR limit.
Along the critical line of a positive chemical potential,
the conformal symmetry is broken for $\aLR < 2$ in the effective action
while the Ising exponent $\beta$ is unchanged in the test of a quantity
that corresponds to the Ising order parameter in the SR limit.
Although there is no rigorous mapping between the Kitaev and
Ising chains at finite $\aLR$, their empirical similarity
motivates us to consider the possibility that the same phenomenon
could occur in the AF-LR-TFIC.

Detecting the breakdown of conformal invariance must go beyond
the test of the central charge. As discussed in Ref.~\cite{Patil2017}, 
the central charge measured from the entanglement entropy is not
a sufficient tool to examine conformal invariance because
the logarithmic system-size scaling behavior of the entanglement entropy
is not exclusive to conformal invariance.
A criterion must be set based on a behavior or quantity fully restricted
by conformal field theory (CFT).
The scaling and functional form of the correlation function
in cylindrical geometry was proposed as a strict indicator \cite{Patil2017},
which we use in this work to examine conformal invariance.

Our VMC+RBM calculations investigate these questions about
criticality and conformal invariance.
First, we examine the contrasting evidence from 
the previous TDVP and DMRG studies \cite{Koffel2012,Vodola2016,Sun2017,Puebla2019}.
In the finite-size scaling (FSS) analysis of the order parameter,
we find that the critical exponents $\nu$, $\beta$, and $\gamma$
are indeed very close to the SR Ising values for our examined range
of $0.5 \le \aLR \le 3$.
In contrast, the correlation function exponent and the central charge
exhibit deviations from the SR Ising values
for $\aLR \lesssim 2$ and $\aLR \lesssim 1$, respectively.
These observations verify the previous results.
However, the change of the criticality is inconclusive 
in these contrasting behaviors of the measured exponents,
and as discussed above, the deviation of the central charge 
is not sufficient to claim the breakdown of conformal invariance.

We thus provide additional tests for the Ising criticality and
the breakdown of conformal invariance, examining the universal Binder ratio
and the CFT description of the correlation function, respectively.
It turns out that the critical Binder ratio \cite{Horita2017} becomes
increasingly different from the universal ratio of the 2D Ising model 
as $\aLR$ decreases below $2$,
indicating the deviation from the criticality of the SR limit.
In the CFT test of the correlation function, we find evidence of
the conformal invariance breakdown from a mismatch between
the CFT description and our measurement,
which becomes very pronounced for $\aLR < 2$. 
However, the slope of the scaled correlation function \cite{Patil2017}
appears to remain small but nonvanishing for $\aLR \ge 2$.
This raises the possibility the breakdown threshold 
of conformal invariance in the AF-LR-TFIC may differ from
the threshold discussed in the Kitaev chain. 

This paper is organized as follows. 
The AF-LR-TFIC model Hamiltonian and the details of
the VMC+RBM calculations are described in Sec.~\ref{sec:method}. 
The previous estimates of the critical exponents
and the central charge are verified through Sec.~\ref{sec:result_exponent},
Sec.~\ref{sec:result_eta}, and Sec.~\ref{sec:result_renyi},
which is followed by our tests using different methods.
In Sec.~\ref{sec:result_binder}, the test for the Ising universality 
using the Binder ratio is given. 
In Sec.~\ref{sec:result_cft}, the CFT test of the correlation 
function is given to identify the conformal invariance breakdown.
Conclusions are given in Sec.~\ref{sec:conclusion}.

\section{Model and VMC+RBM Calculations}
\label{sec:method}

We consider the AF-LR-TFIC Hamiltonian \cite{Koffel2012} given as
\begin{equation} \label{eq:TFIC}
    \hat{H} = \sin\theta \sum_{i < j} J_{ij} \hat{\sigma}^x_i \hat{\sigma}^x_j
    + \cos\theta \sum_i \hat{\sigma}^z_i, 
\end{equation}
where $\theta$ is in the range of $0 < \theta < \pi/2$ for the AF coupling, 
and the site indices $i$ and $j$ run from $1$ to $L$ in the chain
of length $L$.
We impose PBC as the boundary conditions that are necessary
for the test of the CFT description of the correlation function
constructed in a cylindrical space-time geometry.
In the implementation of the algebraically decaying LR interaction
under PBC, we choose to write $J_{ij}$ with a range cutoff that
increases with the system size $L$ by adopting the formulation
used in the LR-Kitaev chain \cite{Vodola2014,Lepori2016} as
\begin{equation}
J_{ij} = 
\begin{cases}
|i-j|^{-\alpha_\mathrm{LR}} & \text{for  } |i-j| < L/2, \\
(L-|i-j|)^{-\alpha_\mathrm{LR}} & \text{for  } |i-j| > L/2.
\end{cases}
\end{equation}

We choose RBM as an ansatz of a trial wave function for VMC simulations
to find an approximate ground state \cite{Carleo2017}.
A trial state can be written as
$|\Psi\rangle = \sum_\mathbf{s} \Psi(\mathbf{s};\mathscr{W})|\mathbf{s}\rangle$
with the visible variables $\mathbf{s} = (s_1, s_2, \ldots ,s_L)$ of RBM,
where $s_i$ indicates $\sigma^x_i$ for the $\hat{\sigma}^x$-basis 
representation of the given Hamiltonian.  
We impose the translation symmetry under PBC to reduce the number
of variational parameters. 
Following the procedures of Ref.~\cite{Carleo2017},
after integrating out the hidden layer,
one can express the RBM wave function as 
\begin{equation} \label{eq:RBM}
    \Psi(\mathbf{s};\mathscr{W}) = 
    e^{a\sum_{j=1}^L s_j} \prod_{m=1}^L \prod_{i=1}^{n_h} 
    \cosh \bigg[ b_i + \sum_{j=1}^L W_{ij} T_m(s_j) \bigg] ,
\end{equation}
where the translation operator $T$ is defined as 
$T_m(s_j) = s_{j+m}$ with periodicity $s_{j+L} = s_j$,
and $n_h$ is the number of filters given for the symmetry.
On a diagram of RBM, one may illustrate the hidden layer with
$N_h = L n_h$ neurons with $L$-fold degeneracy
of the neural variables enforcing the translation invariance.
In Eq.~\eqref{eq:RBM}, there are $(1 + n_h + L n_h)$
RBM parameters of $\mathscr{W} \equiv \{ a, \mathbf{b}, \mathbf{W} \}$ 
to be optimized using the VMC method.
We adopt complex-valued parameters as suggested in Ref.~\cite{Carleo2017}
for better convergence, although the TFIC Hamiltonian is stoquastic \cite{Park2022}. 
We initialize the RBM by setting $a = 0$ and assigning Gaussian
random numbers with zero mean and variance of $1/(Ln_h)$ 
to $\mathbf{b}$ and $\mathbf{W}$.

\begin{figure}
    \includegraphics[width = 0.48\textwidth]{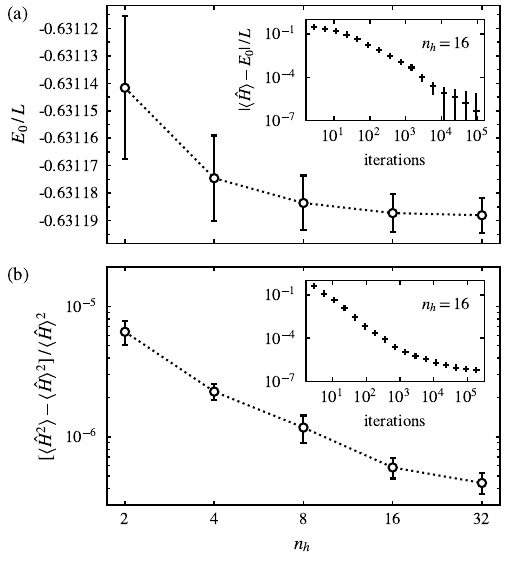}
    \caption{Convergence test of the RBM wave function
    in the VMC search for the ground state. 
    The case with the system of the size $L = 64$ for $\alpha_\mathrm{LR} = 0.5$ is shown for example.
    The estimates of (a) energy density $E_0 / L$ and (b) relative variance
    $(\langle \hat{H}^2 \rangle - \langle \hat{H} \rangle^2) / \langle \hat{H} \rangle^2$
    measured after $2 \times 10^5$ iterations are plotted as a function of $n_h$. 
    The insets show the same quantities for a fixed number of filters $n_h = 16$
    monitored during the iterations of the parameter updates.
    The data points in the insets represent the averages measured
    in the logarithmic bins of iteration numbers.
    The error bars are measured with ten independent RBM wave function samples.
    }
    \label{fig1}
\end{figure}

In VMC calculations, we optimize the RBM parameters using
the stochastic reconfiguration method to construct
the natural gradient \cite{Sorella2001,Sorella2007,Neuscamman2012}.
This method can be described as the imaginary-time evolution of
a trial state, providing a new state projected in the space of
$\{|\Psi\rangle, \partial_1 |\Psi\rangle, \partial_2 |\Psi\rangle, \ldots \}$,
where $\partial_i |\Psi \rangle \equiv \frac{\partial |\Psi\rangle}{\partial \mathscr{W}_i}$.
These procedures propose an update of the variational parameter as
$\mathscr{W}_i^\mathrm{new} = \mathscr{W}_i^\mathrm{old} + \mu \delta\mathscr{W}_i$,
where $\delta\mathscr{W}_i$ is determined by solving
the linear equation $\mathbf{S} \delta\mathscr{W} = - \mathbf{f}$.
The essential numerical procedures are to evaluate
the overlap matrix $\mathbf{S}$ and the force vector $\mathbf{f}$, 
\begin{eqnarray}
    S_{ij} &=& \left\langle \Delta^*_i \Delta_j \right\rangle_\mathrm{mc}
    - \left\langle \Delta^*_i \right\rangle_\mathrm{mc} 
      \left\langle \Delta_j \right\rangle_\mathrm{mc}, \\  
    f_i &=& \left\langle \Delta^*_i E_\mathrm{loc} \right\rangle_\mathrm{mc}
    - \left\langle \Delta^*_i \right\rangle_\mathrm{mc}
      \left\langle E_\mathrm{loc} \right\rangle_\mathrm{mc},
\end{eqnarray}
where the derivative $\Delta_i$ and the local energy $E_\mathrm{loc}$ are
\begin{equation}
\Delta_i \equiv \frac{\partial_i \Psi(\mathbf{s};\mathscr{W})}{\Psi(\mathbf{s};\mathscr{W})}
\quad \text{and} \quad E_\mathrm{loc} \equiv \sum_{\mathbf{s}^\prime}
\langle \mathbf{s} | \hat{H} | \mathbf{s}^\prime \rangle 
\frac{\Psi(\mathbf{s}^\prime;\mathscr{W})}{\Psi(\mathbf{s};\mathscr{W})}.
\end{equation}
The expression $\langle A \rangle_\mathrm{mc} \equiv \sum_\mathbf{s} P(\mathbf{s}) A(\mathbf{s})$
denotes the Monte Carlo (MC) measurement of $A(\mathbf{s})$ with probability
$P(\mathbf{s}) \propto |\Psi(\mathbf{s};\mathscr{W})|^2$.
We use the conjugate gradient algorithm with the Jacobi preconditioner
to solve the linear equation without explicitly storing the $S$ matrix
following the cost-reducing recipe of Ref.~\cite{Neuscamman2012}.
For numerical stability, we use the regularization scheme introduced
in Ref.~\cite{Carleo2017}, where at the $p$th iteration,
$S_{ij}$ is replaced by
$S_{ij} ( 1 +  \lambda_p \delta_{ij})$ 
with $\lambda_p = \max(\lambda_0 b^p, \lambda_\mathrm{min})$. 
We use the parameters $\lambda_0 = 100$, $b = 0.9$, and $\lambda_\mathrm{min} = 0.01$.
The learning rate $\mu$ is initially set to $0.1$
and increased by $0.1$ for every $10000$ iterations
until it becomes unity.

We monitor the convergence of $|\Psi\rangle$ to the ground state 
by evaluating $\langle \hat{H} \rangle$ and the relative variance
defined as
\begin{equation}
    \tilde{\sigma}_E \equiv 
    \frac{\langle \hat{H}^2 \rangle - \langle \hat{H} \rangle^2}{\langle \hat{H} \rangle^2}.
\end{equation}
The relative variance $\tilde{\sigma}_E$ should be precisely zero
when $|\Psi\rangle$ becomes an exact eigenstate.
However, in practice, it does not decrease below a certain value
in VMC simulations. Probable systematic causes may include
the limited expressive power of a finite-size neural network with a finite $n_h$,
despite the universal approximation theorem, and the stochastic
fluctuations in MC measurements that can affect the linear solver.

Figure~\ref{fig1} presents an example of the convergence test
performed at the critical point in the system of size $L=64$
for the LR exponent $\aLR = 0.5$. Convergence tends to slow down as
$\aLR$ decreases in this LR-AF system. At the critical point, it
typically takes about an order of $10^5$ iterations until the energy
and variance become saturated within the scale of their fluctuations
over the iterations. 
We find that the accuracy level indicated by $\tilde{\sigma}_E$
after saturation depends essentially on the number of filters $n_h$.
In our VMC calculations for the ground state, we set the convergence criterion
as $\tilde{\sigma}_E < 10^{-6}$, which, for example, is achieved
for $n_h > 8$ in Fig.~\ref{fig1}. 
In our tests, $n_h = 16$ suffices to satisfy $\tilde{\sigma}_E < 10^{-6}$
for system sizes up to $L = 128$ in the range of $\aLR$ that we consider.

\begin{figure*}
    \includegraphics[width = 0.96\textwidth]{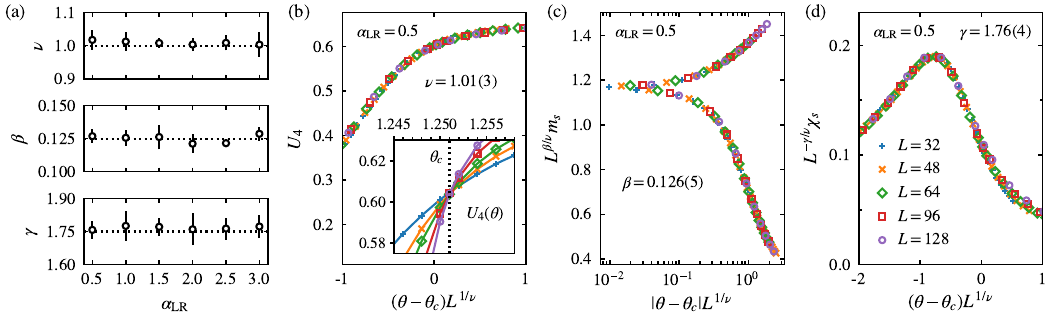}
    \caption{FSS analysis of RBM observables. 
    (a) The estimates of the critical exponents $\nu$, $\beta$, $\gamma$
    are plotted in the range of $\aLR$ between $0.5$ and $3$.
    The dotted lines are given for comparison with the SR Ising values.
    The FSS collapse tests with the critical exponents
    are demonstrated at $\aLR = 0.5$ for the data of
    (b) the Binder's cumulant $U_4$, (c) the AF order parameter $m_s$,
    and (d) the susceptibility $\chi_s$.
    The inset of (b) shows the crossing point of $U_4$ 
    locating the critical point $\theta_c$.
    }
    \label{fig2}
\end{figure*}

\section{Results and Discussions}
\label{sec:result}

Using the RBM wave function $\Psi(\mathbf{s})$
obtained in the VMC optimizations at a given $\theta$, 
we measure the moments of staggered magnetization
including the AF order parameter, the two-point correlation function,
and the second R\'enyi entanglement entropy.
For a given RBM sample, the MC averages are calculated with $4 \times 10^8$
configurations of $\mathbf{s}$ sampled from the probability distribution
$P(\mathbf{s}) \propto |\Psi(\mathbf{s})|^2$ using the Metropolis algorithm. 
We obtain ten RBM wave function samples from independent
VMC calculations. We find that the standard error of the measurement
based on one RBM sample is typically smaller than the sample-to-sample
fluctuations, and thus, we estimate the error bar by the standard deviation
of the measurements over the RBM samples.

In this section, we first present the FSS analysis to estimate
the critical exponents and the central charge
for comparison with the previous TDVP and DMRG results.
Then, we proceed to present our additional tests of
the universal Binder ratio and the CFT description of the correlation function
to examine the Ising criticality and the conformal invariance,
respectively.

\subsection{Order parameter and critical exponents}
\label{sec:result_exponent}

The emergence of the AF order can be detected by measuring
the staggered magnetization in the input layer of the RBM. 
In the AF phase, the operator $\hat{M}_s = \sum_i (-1)^i \hat{\sigma}_i^x$
in each parity sector of the $\mathbb{Z}_2$ symmetry indicates
a finite positive or negative expectation value.
Although our MC sampling does not fix the parity, an alternative quantity
$M_s(\mathbf{s}) = |\sum_i (-1)^i s_i|$ can characterize
the order-disorder phase transition at the level of the RBM wave function. 
We write the order parameter as
\begin{equation}
    m_s = \frac{1}{L} \left\langle M_s \right\rangle_\mathrm{mc}.
\end{equation}
Near a critical point $\theta_c$, the order parameter measured
in a finite system of size $L$ is expected to behave asymptotically as
$m_s(\theta,L) \sim L^{-\beta/\nu}\mathscr{M}^{(\pm)}_\mathrm{o}(|\theta - \theta_c| L^{1/\nu})$
with the critical exponents $\beta$ and $\nu$, where $\mathscr{M}^{(\pm)}_\mathrm{o}$
is a size-independent scaling function.
The corresponding susceptibility can also be defined by the fluctuations
of $M_s$ as
\begin{equation}
    \chi_s = \left\langle M_s^2 \right\rangle_\mathrm{mc} 
    - \left\langle M_s \right\rangle_\mathrm{mc}^2,
\end{equation}
which is expected to follow the FSS ansatz of 
$\chi_s(\theta,L) \sim L^{\gamma/\nu}\mathscr{X}_\mathrm{o}[(\theta - \theta_c) L^{1/\nu}]$
associated with the exponent $\gamma$. 

First we determine the critical point $\theta_c$ for a given $\alpha_\mathrm{LR}$ 
by locating a crossing point of the Binder's fourth-order cumulant, 
\begin{equation}
    U_4 = 1 - \frac{\left\langle M_s^4 \right\rangle_\mathrm{mc}}{3 \left\langle M_s^2\right\rangle_\mathrm{mc}^2},
\end{equation}
between the curves of different $L$'s.
The FSS ansatz of the cumulant is given as
$U_4(\theta, L) \sim \mathscr{U}_\mathrm{o}[(\theta - \theta_c)L^{1/\nu}]$. 
Although $\mathscr{U}_\mathrm{o}$ becomes independent of $L$ 
for a large $L$, a finite-size correction can appear for small $L$'s.
The finite-size correction of the leading order is usually assumed
to be in the form of
$\theta^*_{L,2L} - \theta_c \propto L^{-\tilde{\omega}}$ 
for a crossing point $\theta^*_{L, 2L}$ identified between
two adjacent curves of system sizes $L$ and $2L$.
We determine $\theta_c$ based on this correction-to-scaling ansatz
with the extrapolation to infinite size. 

After locating the critical point $\theta_c$, we estimate the critical
exponents $\nu$, $\beta$, and $\gamma$ by performing the standard
FSS analysis with the FSS ansatz of $m_s$, $\chi_s$, and $U_4$ 
in the critical region.
Figure~\ref{fig2} presents an example of the FSS analysis for $\aLR = 0.5$,
showing that the data points of different $L$'s fall well on
a common scaling curve with our estimates of the critical exponents.
The numerical estimates of the critical exponents and errors are
measured using the \texttt{PYFSSA} package \cite{pyfssa,Melchert2009}. 
We tabulate our estimate of $\theta_c$ and the critical exponents in Table~\ref{tab1}. 
Within the error bars, our estimates of the critical exponents are
very close to the SR Ising values for all the values of $\aLR$ examined
as shown in Fig.~\ref{fig2}(a), 
which is consistent with the previous DMRG results \cite{Sun2017,Puebla2019}.

\subsection{Correlation function exponent}
\label{sec:result_eta}

\begin{figure}
    \includegraphics[width = 0.48\textwidth]{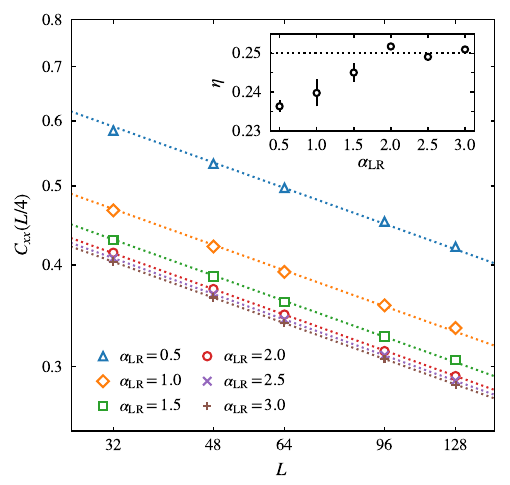}
    \caption{Critical exponent of the spin-spin correlation function.
    The correlation function $C_{xx}(r)$ at $r=L/4$
    is plotted as a function of the system size $L$.
    The inset shows the exponent $\eta$ extracted from 
    the data fitting to $C_{xx}(L/4) \propto L^{-\eta}$. 
    The dotted lines given for comparison indicate
    the SR Ising exponent $\eta = 1/4$.
    }
    \label{fig3}
\end{figure}

We also examine the critical exponent $\eta$ of the correlation function.
At the critical point $\theta_c$ given above in Sec.~\ref{sec:result_exponent},
we measure the spin-spin correlation function,
\begin{equation} \label{eq:cxx}
    C_{xx}(r) =  \langle \hat{\sigma}^x_i \hat{\sigma}^x_{i+r} \rangle 
    =  \langle s_i s_{i+r} \rangle_\mathrm{mc}\,,
\end{equation}
where the distance $r$ runs from $1$ to $L/2$ in the periodic chain,
and this expression is independent of the site index $i$ due to
the translational symmetry imposed in our RBM wave function ansatz.
The asymptotic algebraic decay of $C_{xx}(r) \propto r^{-\eta}$
expected at a critical point can be written for a finite system of length $L$
as $C_{xx}(bL) \propto L^{-\eta}$ at $r = bL$,
which is displayed for our choice of $b = 1/4$ in Fig.~\ref{fig3}.
We extract $\eta$ from the linear fit of the data in the log-log plot.
It turns out that the estimate of $\eta$ is consistently smaller than
the SR Ising value $1/4$ when $\aLR$ decreases below $2$,
implying that the SR Ising criticality does not hold for $\aLR < 2$.
These observations are consistent with the previous
TDVP result \cite{Koffel2012}, where the threshold at $\aLR = 2.25$ 
was suggested.

However, it is worth noting that a change in $\eta$ is connected
to changes in the other exponents through the hyperscaling relations
$\gamma / \nu = 2 - \eta$ and $2\beta/\nu = d + z - 2 + \eta$.
The latter implies an apparent conflict between
the previous TDVP and DMRG results \cite{Koffel2012,Sun2017,Puebla2019}.
The TDVP estimate of $\eta$ decreases from $1/4$ for $\aLR \lesssim 2$,
whereas the DMRG studies reported 
$\nu \simeq 1$, $\beta \simeq 1/8$, and $z \simeq 1$.
We have verified some of these critical exponents, but 
we cannot reconcile this conflict with the limited accuracy
of our calculations. Thus, the exponents at the present accuracy may be
unreliable to determine whether the Ising criticality of the SR limit
survives or breaks down at a finite $\aLR$.
This emphasizes the need for a test that does not rely
on the estimates of the exponents.

\begin{table}[b]
    \begin{ruledtabular}
    \begin{tabular}{ccccccc}
        $\alpha_\mathrm{LR}$ & $\theta_c$ & $\nu$ & $\beta$ & $\gamma$ & $\eta$ & $c_\infty$ \\
        \hline
        3.0 & 0.8714(7) & 1.00(4) & 0.128(5) & 1.77(5) & 0.2510(4) & 0.496(5) \\ 
        2.5 & 0.9041(6) & 1.01(2) & 0.122(3) & 1.76(5) & 0.2491(2) & 0.500(4) \\
        2.0 & 0.9489(7) & 1.00(2) & 0.121(7) & 1.76(7) & 0.2518(7) & 0.502(5) \\
        1.5 & 1.012(1) & 1.00(1) & 0.126(9) & 1.77(4) & 0.2450(24) & 0.508(4) \\
        1.0 & 1.103(1) & 1.01(3) & 0.126(6) & 1.78(7) & 0.2398(35) & 0.490(5) \\
        0.5 & 1.251(1) & 1.01(3) & 0.127(5) & 1.76(4) & 0.2363(15) & 0.454(8) \\
    \end{tabular}
    \end{ruledtabular}
    \caption{List of the critical points and exponents. 
    Critical exponents $\nu$, $\beta$, and $\gamma$ are determined
    in the FSS analysis of the collapse of the scaling curve. 
    The exponent $\eta$ is measured from the scaling of
    the spin-spin correlation function along a fixed $r/L = 1/4$
    at the critical point $\theta_c$. The central charge $c_\infty$
    is extracted from the logarithmic scaling of the second R\'enyi entropy.
    }
    \label{tab1}
\end{table}

\subsection{Second R\'enyi entropy and central charge}
\label{sec:result_renyi}

The logarithmic system-size scaling of the entanglement entropy 
at a critical point in one dimension is a useful universal property to measure
the central charge of the CFT that characterizes the phase transition
\cite{Vidal2003,Calabrese2004,Calabrese2009}. 
In the previous estimate of the central charge using
the TDVP \cite{Koffel2012}, DMRG \cite{Vodola2016},
and generalized Hatree-Fock \cite{Kaicher2023} methods, 
the von Neumann entropy was examined under
open boundary conditions (OBC). Instead, we consider the second
R\'enyi entropy for the measurement using the RBM wave function under PBC.  
For the bipartition of a system into subsystems $A$ and $B$, 
the R\'enyi entropy of an order $n$ for $\rho_A$ is written as
\begin{equation}
    S_n (\rho_A) = \frac{1}{1-n} \ln \mathrm{tr} \rho_A^n,  
\end{equation}
where $\rho_A \equiv \mathrm{tr}_B \rho$ is the reduced density 
matrix of $A$ for a pure state $\rho$. The von Neumann entropy
is recovered at the limit of $n=1$.
For the universality class fixed by the CFT, 
the von Neumann and R\'enyi entropies at the critical point
indicate the same central charge $c$ in the leading-order FSS
behavior. For PBC, the asymptotic scaling behavior of
$S_n$ \cite{Calabrese2009} for half-chain bipartition
is written as
\begin{equation}
    S_n = \frac{c}{6}\left( 1 + \frac{1}{n} \right) \ln L + c^\prime_n,
\end{equation}
where $c^\prime_n$ is a nonuniversal constant.

\begin{figure}
    \includegraphics[width = 0.48\textwidth]{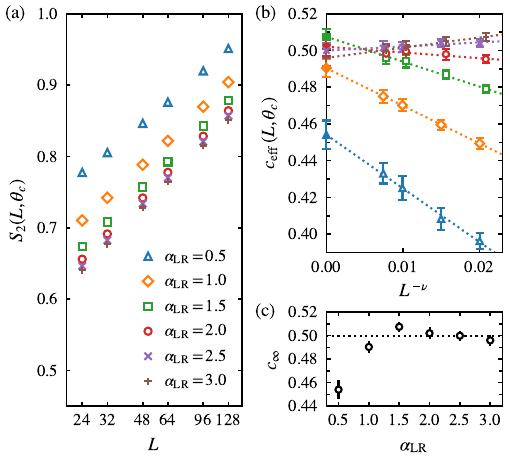}
    \caption{Estimate of the central charge. 
    (a) The second R\'enyi entropy $S_2$ of a half chain 
    is plotted at the critical point $\theta_c$ as a function of system size $L$.
    (b) The effective central charge $c_\mathrm{eff}$ (empty symbols) is plotted
    as a function of $1/L$. Solid symbols and their error bars indicate 
    our estimate of the central charge $c_\infty$ and its standard error obtained
    from the linear fit (dotted lines) of the $c_\mathrm{eff}$ data.
    In (c), $c_\infty$ is plotted as a function of $\aLR$.
    }
    \label{fig4}
\end{figure}

The second R\'enyi entropy $S_2$ can be reliably measured in
QMC calculations by using the replica trick \cite{Hastings2010}, 
which has been successfully applied to the VMC calculations
with the RBM wave function \cite{Torlai2018}.
We consider only $S_2$, but a method was proposed to compute $S_n$
of the higher $n$ and to approximate $S_1$
in a different NQS representation \cite{Wang2020}. 
Measuring $S_2$ requires two copies of the RBM state, 
namely $\mathbf{s}^{(1)}$ and $\mathbf{s}^{(2)}$, 
sampled from the joint probability distribution
$P(\mathbf{s}^{(1)},\mathbf{s}^{(2)}) \propto |\Psi(\mathbf{s}^{(1)})|^2 |\Psi(\mathbf{s}^{(2)})|^2$. 
Each copy can be rewritten in a bipartite basis of
$\mathbf{s} \equiv (\mathbf{s}_A, \mathbf{s}_B)$, where
$\mathbf{s}_A$ and $\mathbf{s}_B$ are associated with the subsystems $A$ and $B$. 
Then, one can obtain $e^{-S_2}$ by measuring the swapping operator on $A$ as
\begin{equation}
     e^{-S_2} = 
    \left\langle 
    \frac{\Psi(\mathbf{s}^{(2)}_A, \mathbf{s}^{(1)}_B)}{\Psi(\mathbf{s}^{(1)}_A, \mathbf{s}^{(1)}_B)} 
    \frac{\Psi(\mathbf{s}^{(1)}_A, \mathbf{s}^{(2)}_B)}{\Psi(\mathbf{s}^{(2)}_A, \mathbf{s}^{(2)}_B)} 
    \right\rangle_\mathrm{mc}.
\end{equation}

Figure~\ref{fig4}(a) shows the measurements of $S_2$ for different $L$'s
at the critical point. While the expected asymptotic behavior of
$S_2(L) = \frac{c}{4} \ln L + c'_2$ is apparent,
a finite-size correction that decays with increasing $L$ may exist.
To estimate $c$ from the $S_2$ data of finite $L$'s, 
we define the effective central charge as
\begin{equation} \label{eq:c_eff}
    c_\mathrm{eff}(L) = \frac{4}{\ln 2} [S_2(L) - S_2(L/2)],
\end{equation} 
where the central charge can be estimated by the extrapolation
of $c_\mathrm{eff}(L)$ to $L=\infty$ along a model curve of the finite-size behavior.
According to the previous FSS analysis at a second-order phase transition \cite{Campostrini2014},
the finite-size correction of $S_2$ would be proportional to $L^{-\nu}$ under PBC.
Since Eq.~\eqref{eq:c_eff} inherits the correction of $S_2$,
we fit the data to the line of $c_\mathrm{eff}(L) = c_\infty + a L^{-\nu}$
as shown in Fig.~\ref{fig4}(b), where
the fitting parameter $c_\infty$ is our estimate of the central charge.

In Fig.~\ref{fig4}(c), the estimate of $c_\infty$ for $\aLR \lesssim 1$
indicate a deviation from $c=1/2$ of the 2D Ising CFT.
While such a deviation in the central charge
was also previously observed for a similar range of $\aLR$
\cite{Koffel2012,Vodola2016},
it is worth noting that the direction of a deviation appears
to vary depending on the types of entanglement entropy or boundary conditions.
In the present work based on $S_2$ under PBC, we observe the decrease
of $c_\infty$ that goes below $1/2$ for a small $\aLR$.
In contrast, the use of $S_1$ under OBC in the previous studies showed
a large increase in central charge as $\aLR$ decreases below $1$.

So far, we have verified the previous TDVP and DMRG
results on the critical exponents and the central charge
\cite{Koffel2012,Vodola2016,Sun2017,Puebla2019}.
However, questions remain about the Ising criticality and
the conformal invariance. As discussed through the hyperscaling relations,
we still need to go beyond the estimate of the critical exponents
to see whether the 2D Ising universality class genuinely
holds for any finite $\aLR$ or if there is a criterion for
how large $\aLR$ should be to retain the criticality of the SR limit.
On the other hand, the deviation of the central charge does not provide
a sufficient condition to identify the breakdown of
conformal invariance \cite{Patil2017}.
In the following sections, we thus provide additional stringent tests
of the criticality change and the conformal invariance breakdown.

\subsection{Critical Binder ratio}
\label{sec:result_binder}

\begin{figure}
    \includegraphics[width = 0.48\textwidth]{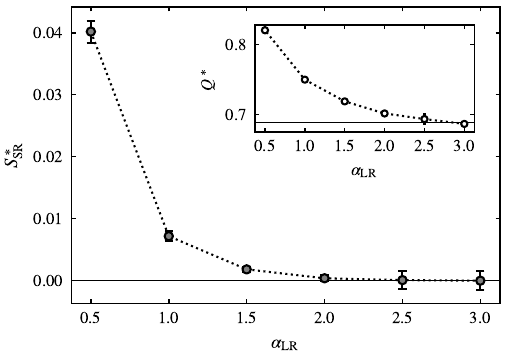}
    \caption{Test of the critical Binder ratio. The self-combined
    ratio $S_\mathrm{SR}^*$ and the Binder ratio $Q^*$ at the critical point 
    are plotted as a function of $\aLR$. The data points are extrapolated
    to infinite size.
    The horizontal solid lines indicate the SR limit.
    }
    \label{fig5}
\end{figure}

For an alternative test for the Ising criticality, we consider the Binder ratio,
$Q \equiv \langle M_s^2 \rangle^2_\mathrm{mc} / \langle M_s^4 \rangle_\mathrm{mc}$,
of the second and fourth moments of the staggered magnetization.
The Binder ratio at a critical point exhibits a particular value
contributing to the universality of the critical behavior,
while the value depends on the boundary conditions and
the aspect ratio of the system
(see, for instance, Refs.~\cite{PrivmanBook,Selke2007} and references therein).
The critical Binder ratio has been used as a reliable ingredient
to identify the universality class in the classical long-range 
Ising model \cite{Horita2017}, which inspires us
to perform the same test of how the Binder ratio depends
on $\aLR$ for the critical RBM wave function in the AF-LR-TFIC.

\begin{figure*}
    \includegraphics[width = 0.96\textwidth]{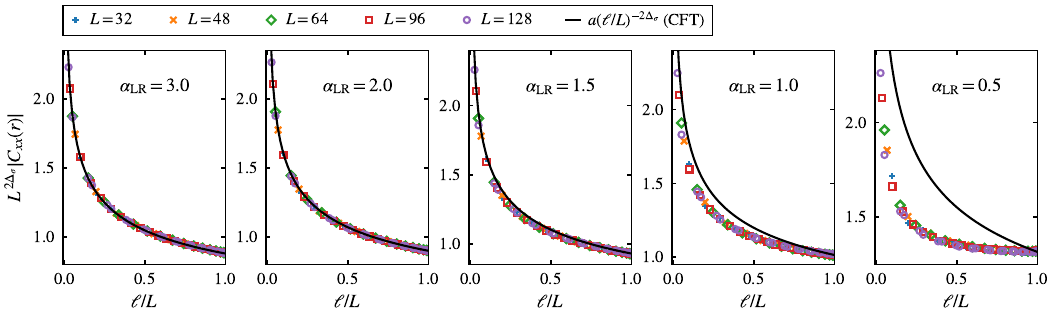}
    \caption{FSS analysis of the spin-spin correlation 
    function. The data collapse of $L^{2\Delta_\sigma}|C_{xx}(r)|$
    is examined as a function of the scaled chord length $\ell/L \equiv \sin(\pi r/L)$ 
    with the measured exponent $\eta = 2\Delta_\sigma$.
    The solid line indicates the CFT-predicted form of $a(\ell/L)^{-2\Delta_\sigma}$
    given for comparison with the scaled curve of the measured correlation function.
    }
    \label{fig6}
\end{figure*}

In the SR limit, at the exact critical point $\theta_c = \pi/4$,
we obtain the value of $Q_\mathrm{SR}^* = 0.689(4)$
from the power-law extrapolation of $Q(L)$ to infinite $L$. 
This particular value of the ratio has not previously been known
for the AF-TFIC, but it turns out that the corresponding value of
the cumulant $U_4^* = 0.516(3)$ is in good agreement with
the previous MC estimate of $U_4^* = 0.514(1)$
reported in the classical 2D Ising model subject to the mixed boundary
conditions where the system is periodic in one direction
and open in the other direction \cite{Selke2006}.
The implicit connection between the mixed boundary conditions and
the cylindrical geometry of our periodic chain
under the imaginary-time evolution at zero temperature may expect
the universal value of the Binder ratio in the SR limit.

For a finite $\aLR$, we consider the indicator called
the self-combined Binder ratio proposed in Ref.~\cite{Horita2017},
\begin{equation}
    S_\mathrm{SR}(L) = \frac{1}{Q_\mathrm{SR}^*} Q(L) 
    + \frac{1}{Q(L)} Q_\mathrm{SR}^* - 2,
\end{equation}
which removes the leading-order finite-size correction in $Q(L)$
and thus exhibits better convergence with increasing $L$ if an 
accurate value of $Q_\mathrm{SR}^*$ is provided.
Figure~\ref{fig5} displays the value of
$S_\mathrm{SR}^* \equiv \lim_{L\to\infty} S_\mathrm{LR}(L)$
obtained from the power-law extrapolation to infinite $L$. 
It turns out that while $S^*_\mathrm{SR}$ is almost zero for $\aLR = 3$
and $2.5$, the deviation of $S^*_\mathrm{SR}$ appears
for $\aLR \lesssim 2$ and increases as $\aLR$ decreases. 
The estimate of $Q^* \equiv \lim_{L\to\infty} Q(L)$ shows a similar increase
from the value of the SR limit as $\aLR$ decreases, although it still indicates
a slight deviation even for $\aLR = 2.5$ and $3$ where $S^*_\mathrm{SR} \simeq 0$.
This is consistent with the observation in Ref.~\cite{Horita2017},
verifying that $S_\mathrm{LR}(L)$ converges better at a finite $L$.
Our data suggest that the threshold for the SR Ising criticality
is possibly around $\aLR = 2$ above which $S_\mathrm{SR}^*$ is
zero within the error bars.

\subsection{CFT test of the correlation function}
\label{sec:result_cft}

We detect the breakdown of conformal invariance by identifying
a mismatch between the measured correlation function and the CFT description,
following the strategy of Ref.~\cite{Patil2017}.
The CFT in a cylindrical space-time geometry restricts
the scaling and functional form of the correlation functions \cite{HenkelBook,FrancescoBook}. 
In the presence of conformal invariance at a critical point, the spin-spin
correlation function in Eq.~\eqref{eq:cxx} must behave asymptotically as
\begin{equation} \label{eq:cxx_CFT}
    C_{xx}(r) \propto \ell^{-2\Delta_\sigma} 
    =  \left[ L\sin\left(\pi \frac{r}{L}\right) \right]^{-2\Delta_\sigma},
\end{equation}
where the scaling variable $\ell \equiv L\sin(\pi r/L)$ is the chord length,
and the scaling dimension $\Delta_\sigma$ corresponds to
a half of the decay exponent $\eta$. 
The test of Eq.~\eqref{eq:cxx_CFT} requires the estimate of the exponent
$\eta = 2\Delta_\sigma$, which we have already measured in the FSS analysis of
$C_{xx}(L/4) \propto L^{-\eta}$ along $r = L/4$ for different $L$'s
in Sec.~\ref{sec:result_eta}.  
Using the estimate of $\eta$ obtained for each $\aLR$, 
we present how the measured $C_{xx}$ deviates
from the CFT form of Eq.~\eqref{eq:cxx_CFT} as $\aLR$ changes. 

Figure~\ref{fig6} shows a good FSS collapse of the data points of
$L^{2\Delta_\sigma}|C_{xx}(r)|$ on a common scaling curve
with the measured value of $\eta = 2\Delta_\sigma$, which is plotted
as a function of $\ell /L$.
This allows us to make a graphical comparison with the CFT curve of
$L^{2\Delta_\sigma}|C_{xx}(r)| \propto (\ell / L)^{-2\Delta_\sigma}$. 
We find that the deviation between the data and the CFT curve becomes
pronounced for $\aLR < 2$, indicating the breakdown
of conformal invariance. This observation also implies that 
the central charge is indeed unreliable in detecting the breakdown
of conformal invariance because the deviation from
the CFT curve already occurs in the range of $\aLR > 1$
where the central charge is found to be very close to $1/2$.

\begin{figure}
    \includegraphics[width = 0.48\textwidth]{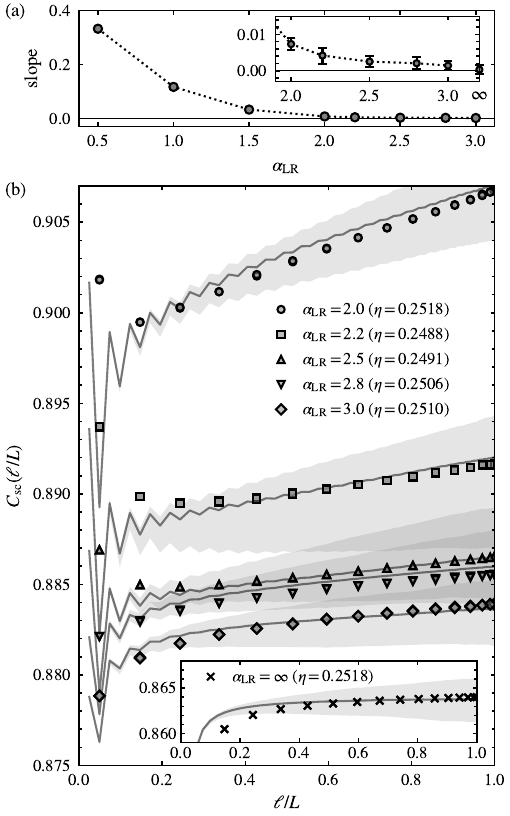}
    \caption{The CFT test of the scaled correlation function.
    Equation~\eqref{eq:csc_CFT} is examined using ten RBM wave function samples
    obtained from independent VMC optimizations at a critical point. 
    (a) The slope of $C_\mathrm{sc}(\ell / L)$ is measured from the straight-line fit 
    to the data of $L = 128$ for $\ell / L > 0.6$.  
    (b) The symbols indicate the data of $L=64$ where the sample-to-sample fluctuations
    are smaller than the symbol size.
    The solid line is the average over the RBM samples for $L=128$.
    The shade fills the area between the minimum and maximum magnitudes of the data 
    in the RBM samples for $L=128$.
    }
    \label{fig7}
\end{figure}

A more quantitative CFT indicator can be provided by the scaled
correlation function \cite{Patil2017},
which is the ratio between the measured correlation function
and the CFT prediction,
\begin{equation} \label{eq:csc_CFT}
    C_\mathrm{sc}(\ell/L) = \left[ L\sin\left( \pi \frac{r}{L} \right) \right]^{2\Delta_\sigma} 
    \left| C_{xx}(r) \right|.
\end{equation}
In Fig.~\ref{fig7}, we specify the measured value of $\eta = 2\Delta_\sigma$ 
for each $\aLR$ examined.
The appearance of a nonzero slope tail in the scaled correlation
function $C_\mathrm{sc}$ signals the breakdown of conformal invariance.
Our data shows that $C_\mathrm{sc}(\ell / L)$
exhibits a tail that is almost linear in $\ell / L$, which allows us
to measure the slope from a straight-line fit of the tail part. 

Figure~\ref{fig7} shows a notable increase in slope as $\aLR$
decreases below $2$, which justifies the graphical deviation
of the data from the CFT curve in Fig.~\ref{fig6}. However,
for $2 \le \aLR \le 3$, where the data and the CFT curve appear
to match in Fig.~\ref{fig6}, it turns out that the slope decreases
with increasing $\aLR$ but remains finite in the range of $\aLR$ examined.
This can be compared with the benchmark case of the SR limit,
where the slope is near zero within the error bar.
Although it is challenging to locate the threshold of the precise zero slope
within the limited accuracy of our calculations,
the nonvanishing slope raises the possibility that the breakdown threshold
in the AF-LR-TFIC
may differ from the threshold in the Kitaev chain,
where the conformal symmetry was argued to be broken at $\aLR = 2$
in the approximate renormalization group approach \cite{Lepori2016}.

Accurate correlation function calculations on the power-law decaying tail
are crucial to determine the threshold. However, apart from  
the common finite-size issues, our RBM wave function has
its limitations in accuracy.
The benchmark in the SR limit indicates the measured exponent $\eta = 0.2518(6)$ 
that is slightly deviated from the exact value $1/4$,
implying a bias in the optimized wave function.
Although increasing the number of filters $n_h$ can enhance
the expressivity and hence the accuracy of the RBM ansatz,
our implementation is practically limited to $n_h = 16$
due to the computational time costs for the FSS analysis.
Moreover, because of the same time costs, only a few samples of 
the optimized RBM wave function are generated to measure fluctuations
across independent VMC optimizations.
These sample-to-sample fluctuations tend to increase with the system size
and cause a significant uncertainty in the scaled correlation function
for $L = 128$, as shown in Fig.~\ref{fig7}(b).
These practical limitations pose numerical challenges
for our VMC+RBM simulations to find the precise breakdown
threshold of conformal invariance.

\section{Summary and Conclusions}
\label{sec:conclusion}

We have investigated criticality and conformal invariance
at a quantum phase transition in the AF-LR-TFIC using the VMC calculations
with the RBM trial wave function ansatz.
Our main findings are from the tests of the universal Binder ratio \cite{Horita2017}
and the CFT description of the spin-spin correlation function \cite{Patil2017}.
The critical Binder ratio exhibits an increasing deviation
from the universal ratio of the SR limit when $\aLR$ decreases below $2$,
implying that the criticality for $\aLR < 2$ is different from the 2D Ising class.
On the other hand, in the test of the correlation function,
we found evidence of the conformal invariance breakdown from the deviation
between the form of the correlation function and the CFT description.
The deviation from the CFT description becomes more pronounced as $\aLR$ decreases,
although the precise threshold of the breakdown is yet to be determined.

These findings present progress in characterizing the phase transition
in the AF-LR-TFIC beyond the observations of the critical exponents
and the central charge.
In the FSS analysis to extract the critical exponents,
we observed that the exponents $\nu$, $\beta$, and $\gamma$ are very close to
the SR Ising exponents for the examined range of $0.5 \le \aLR \le 3$.
In contrast, the decay exponent $\eta$ of the correlation function and
the central charge extracted from the second R\'enyi entropy differ from
the SR Ising values when $\aLR$ becomes small enough.
Although these observations are consistent with the previous TDVP and DMRG
calculations \cite{Koffel2012,Vodola2016,Sun2017,Puebla2019},
the central charge is insufficient to diagnose the conformal invariance
breakdown \cite{Patil2017},
and the critical exponent estimates in the present accuracy are
inconclusive on the change of the criticality.

Our VMC+RBM calculations have demonstrated the practical applicability
of the NQS framework for studying quantum phase transitions.
However, there are also unsolved issues in our calculations
that need further numerical advancement.
Finding the precise breakdown threshold of conformal invariance
needs high-precision calculations of the correlation function or
more sensitive indicators of conformal symmetry,
such as Klein bottle entropy \cite{Tu2017}.
In addition, the apparent mismatches between the correlation function
exponent and the other critical exponents need careful investigation
to examine their hyperscaling relations with higher numerical accuracy.

\begin{acknowledgements}
J.K. and D.K. contributed equally to this work.
We thank Synge Todo and Hong-Hao Tu for the fruitful discussions
during the ASG meeting at the PCS-IBS. 
This work was supported by the Basic Science Research Program
through the National Research Foundation of Korea (NRF-2019R1F1A106321) 
and also by the KIAS associate member program. 
Computing resources are provided by 
the KISTI supercomputing center (KSC-2021-CRE-0165). 
We appreciate APCTP and PCS-IBS for their hospitality
during the completion of this work. 
\end{acknowledgements}

\bibliography{paper}

\end{document}